\documentclass[preprint,prd,aps,showpacs,showkeys,nofootinbib]{revtex4}
\usepackage{graphicx}
\usepackage{dcolumn}
\usepackage{bm}
\begin{document}
\title{The one-loop on-shell renormalization of some vertexes in MSSM}
\author{ZHAO Shu-Min$^{1;1)}$\email{zhaosm@mail.hbu.edu.cn}%
\quad FENG Tai-Fu$^{1}$\quad WANG Fang$^{2}$
\quad GAO Tie-Jun$^{3}$ \quad ZHANG Yin-Jie$^{1}$}
\affiliation{1 Department of Physics, Hebei University, Baoding 071002,
China\\2 Department of Electronic and Information Engineering, Hebei University, Baoding
071002, China\\
3 Department of Physics, Dalian University of Technology, Dalian 116024, China}

\begin{abstract}
The on-shell renormalization scheme for the electroweak theory is well
studied in the standard model(SM), but a consistent on-shell renormalization scheme for the
minimal supersymmetric standard model(MSSM) is still unknown. In MSSM,
we study the on-shell scheme for three
vertexes$Z\overline{l^I}l^I , W^+\overline{\nu^I}l^I $ and $\tilde{L}^*_i\overline{\chi^0_{\alpha}}l^I$ with
virtual SUSY particles (chargino, sneutrino, neutralino and slepton) at one-loop order. Instead of amplitude of a single triangle diagram, the sum of amplitude of triangle diagrams belonging
to one suit can be renormalized in the on-shell scheme. One suit points out that the internal virtual particles are consistent. Zero-momentum scheme is also used
for the renormalization. The two schemes can make the renormalized results decoupled. In MSSM, some special characters
 of the on-shell scheme are shown here. This work is propitious to complete the on-shell renormalization scheme in MSSM.

\end{abstract}


\maketitle

\section{Introduction}
\indent\indent

As we all know, the quantum field theory is perturbative theory.
That is to say, it can not be solved exactly. To obtain finite
results, renormalization is necessary and there are some typical
renormalization schemes such as MS, $\overline{MS}$, MOM,
zero-momentum and on-shell schemes\cite{MS,MSbar,MOM,onshell1,onshell2}.
In MS scheme, the counter term is just the pole term(${1/\epsilon}, \epsilon\rightarrow 0$).
The counter term is proportional to ${1/\epsilon}+\ln4\pi-\gamma_E$ in $\overline{MS}$ scheme.
The two foregoing schemes have nothing to do with mass. For the on-shell scheme, the renormalization
constants are all obtained under on-shell condition. It is the only physical scheme.
For electroweak theory, the on-shell scheme is the most appropriate one.

If we can resolve the theory accurately, different renormalization schemes can give the same
finite result of any physical process though the functions of the
renormalized parameters are different. However, different physical
predictions are produced from different renormalization schemes and
different renormalized parameters, because of the curtate
perturbation theory.

To obtain the counter term for the UV-divergent diagram, one can take all the external
momenta of the diagram as zero, which is called zero-momentum renormalization scheme.
The advantage is that in arbitrary model each divergent diagram is easy to be renormalized,
and the renormalized results are decoupled\cite{decoupling1,decoupling2}.

We focus on the on-shell renormalization scheme that is popular in
electroweak theory. In the on-shell scheme, the fine structure constant $\alpha$ is
an expansion parameter and defined in the Thomson limit. At any
order of perturbation theory, the physical parameters are the same
as the finite renormalized parameters. They represent clear physical
meaning and can be measured directly in experiment.
The renormalization procedure is summarized in
the counter term approach\cite{onshell1}.

Extending SM, physicists have developed many new models\cite{ewaiwei,susy}
to explain the experimental phenomena. MSSM\cite{MSSM} is the most
attractive one. A lot of experimentalists
of high energy physics are focusing on searching for Higgs bosons in
MSSM. The  colliders (LHC, $e^+e^-$ linear
collider, etc.) will provide abundant information of new physics
beyond SM. In MSSM, the decays $h^0 (H^0,A^0)\rightarrow
\tilde{\chi}^0_m\tilde{\chi}^0_n$, $\tilde{\chi}^0_m\rightarrow h^0
(H^0,A^0)+\tilde{\chi}^0_n(m,n=1,2,3,4)$ and $\check{b}_a\rightarrow
\chi^-_it(a,i=1,2)$ are studied at one-loop order with the on-shell
renormalization scheme\cite{onelsusy,oneloop}, but they do not give analytic
results to show the elimination of UV-divergence apparently.
Considering the one-loop contributions, the
authors\cite{higgs} completed systematic on-shell
renormalization programme for gauge boson and Higgs parts. Radiative
one-loop corrections to the process $e^+e^-\rightarrow
l^+l^-$(hadrons) are calculated with the same
scheme\cite{eetoll}.

For supersymmetric gauge theories, a consistent regularization scheme preserving supersymmetry and gauge invariance is still not known.
Two equivalent ways to solve the problem are shown here.
One is to use an invariant scheme to keep the symmetries to
manifest, where only those counterterms are necessary for renormalization
that they themselves preserve the symmetries.
The other is to use a non-invariant scheme, through using appropriate non-invariant counterterms
to compensate the corresponding symmetry breaking.
With appropriate non-invariant counterterms, W.Hollik\cite{susyQED} shows supersymmetric QED can keep the supersymmetry.
Their study can be generalized to supersymmetric models with soft breakings
and eventually to the supersymmetric extensions of the standard model.
Although the corresponding Slavnov Taylor identities are more involved since they have to express not only
the symmetries but also the spontaneous or soft breaking, their structure is
the same as in SQED. Therefore, this method can also extend to full EW theory of the MSSM.

With the extension of the on-shell scheme of SM, the vertexes $(Z\overline{l^I}l^I ,
W^+\overline{\nu^I}l^I )$ and $\tilde{L}^*_i\overline{\chi^0_{\alpha}}l^I$ are studied at one-loop
order in this work. We find some special characters for the
on-shell scheme in MSSM. Compared with the zero-momentum
scheme, it is easy to find the renormalized results in the on-shell scheme
are decoupled. These selected vertexes are ordinary, and can represent
the general vertexes in MSSM. The study of the on-shell scheme
for these vertexes is propitious to complete the on-shell
renormalization programme of MSSM. If one studies the on-shell
renormalization scheme in other models, it is also helpful.

After the introduction, in Sect.2 we study both the zero-momentum scheme and the on-shell
scheme of two SM vertexes in MSSM. The corresponding results of the SUSY vertex
are shown in Sect.3. In Sect.4, the decoupling behaviors for the counter terms in both renormalization schemes
are researched. Sect.5 is devoted to discussion and conclusion.

\section{Renormalization of SM vertex $(Z\overline{l^I}l^I ,
W^+\overline{\nu^I}l^I )$ in MSSM}
The authors\cite{onshell1,onshell2} studied the on-shell renormalization scheme of electroweak theory
in SM successfully and
completely. Extending the model from SM to MSSM, the condition becomes complex and faint, which needs
more researches. In Feynman gauge, applying both the on-shell and zero-momentum schemes, we study the
two SM vertexes $(Z\overline{l^I}l^I, W^+\overline{\nu^I}l^I )$ with virtual particles
($\tilde{L},\tilde{\chi}^0,\tilde{\nu},\tilde{\chi}^{\pm}$) in this section. The studied one loop
diagrams are shown in Fig.1.
In order to obtain the counter terms, we adopt the naive
dimensional regularization with the anticommuting $\gamma_{_5}$
scheme, where there is no distinction between the first 4 dimensions
and the remaining $D-4$ dimensions\cite{feng1,feng2}.
\vspace{-1.0cm}
\begin{center}
\scalebox{0.7}
{\hspace{-1cm}\includegraphics[width=1\textwidth]{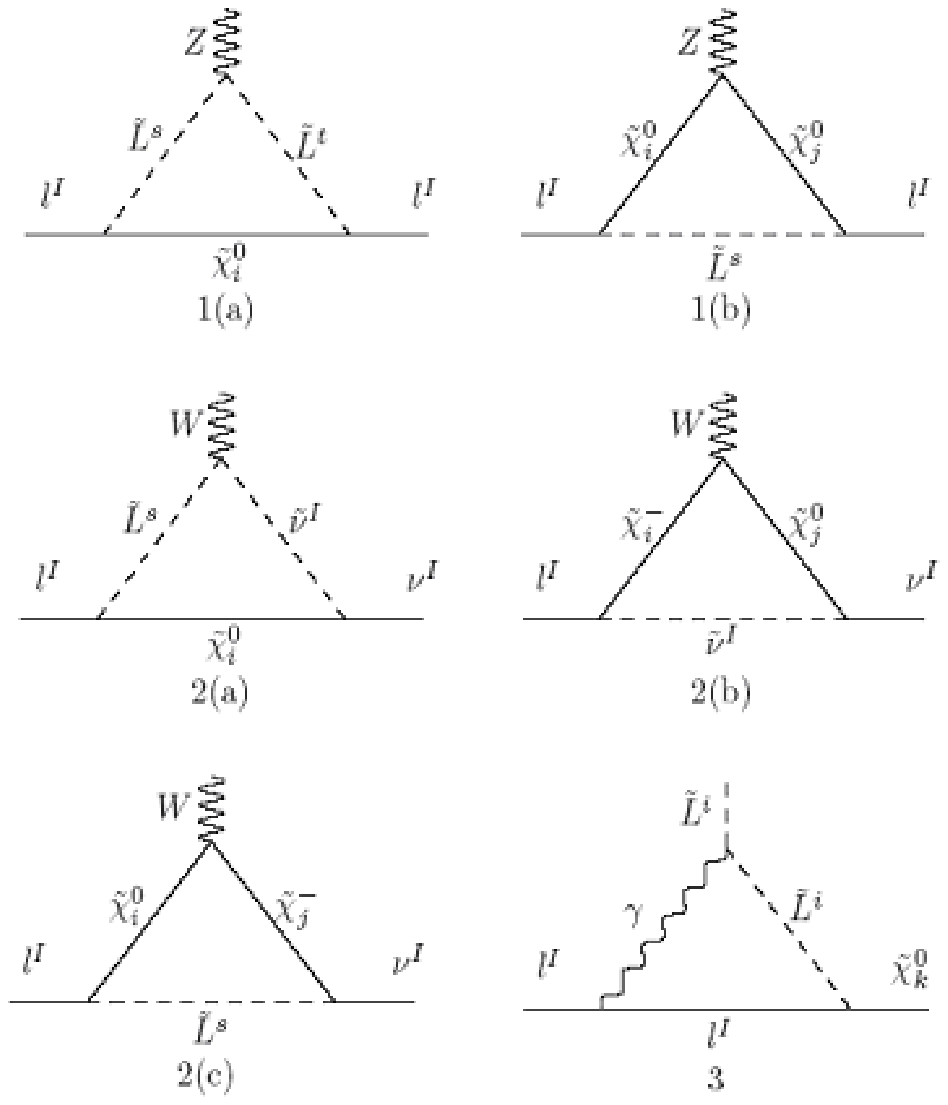}}
\end{center}
\vspace{-8.6cm}
{ \hspace{1.6cm}\footnotesize Figure 1: The studied one loop diagrams.}

\subsection{$Z\overline{l^I}l^I$ vertex with virtual SUSY particles ($\tilde{L},\tilde{\chi}^0)$}
There are two triangle diagrams for $Z\overline{l^I}l^I$ vertex with
virtual SUSY particles $(\tilde{L},\tilde{\chi}^0)$, and they are shown as diagrams 1(a) and 1(b).
The two diagrams are not complete and the results do not satisfy the gauge invariant rule, but
they belong to one suit and can be renormalized with some renormalization constants.
In the zero-momentum scheme, each diagram has its own counter term, and the corresponding renormalized
result is decoupled. Here we show the sum of the counter terms for the two triangle diagrams in the zero-momentum scheme.
\begin{eqnarray}
&&\hspace{-0.5cm}\delta V_{Z\overline{l^I} l^I}^{(ZM),\mu}=\frac{e^3}{64\pi^2s_{_W}c_{_W}}
\Big\{\frac{1-2s_{_W}^2}{2s_{_W}^2}\Big[\Big(\frac{m_{l^I}}{c_{\beta}
m_{_W}}\Big)^2\nonumber\\&&\hspace{-0.5cm}+\frac{1}{c_{_W}^2}\Big]\gamma^{\mu}\omega_-
-\Big[\frac{4s_{_W}^2}{c_{_W}^2} +\Big(\frac{m_{l^I}}{c_{\beta}
m_{_W}}\Big)^2
\Big]\gamma^{\mu}\omega_+\Big\}\Delta_{UV}\nonumber\\&&\hspace{-0.5cm}
+\Big\{\frac{e^3}{256\pi^2s^3_{_W}c_{_W}}
\Big(\frac{1-2s_{_W}^2}{c_{_W}^2}-(1+2s_{_W}^2)(\frac{m_{l^I}}{c_{\beta}m_{_W}})^2\Big)
\nonumber\\&&\hspace{-0.5cm}+\sum_{i,\beta=1}^6\sum_{j=1}^4
\frac{e^3}{4s^3_{_W}c_{_W}}(\mathcal{G})_{i\beta}
(\mathcal{D}^I)_{ij}(\mathcal{D}^I)^*_{\beta j}F_{1}(x_{\tilde{L}_{i}},x_{\tilde{L}_{\beta}},x_{\tilde{\chi}_j^0})
\nonumber\\&&\hspace{-0.5cm}
-\frac{e^3}{8s^3_{_W}c_{_W}}\sum_{s=1}^6\sum_{i,j=1}^4
(\mathcal{D}^I)^*_{si}(\mathcal{R}^*)_{ji}(\mathcal{D}^I)_{sj}
F_{1}(x_{\tilde{\chi}_{i}^0},x_{\tilde{L}_{s}},x_{\tilde{\chi}_j^0})
\nonumber\\&&\hspace{-0.5cm}+\frac{e^3}{4s^3_{_W}c_{_W}}\sum_{s=1}^6\sum_{i,j=1}^4(\mathcal{D}^I)^*_{si}(\mathcal{R})_{ji}
(\mathcal{D}^I)_{sj}\sqrt{x_{\tilde{\chi}_{i}^0}x_{\tilde{\chi}_{j}^0}}
F_2(x_{\tilde{\chi}_{i}^0},x_{\tilde{L}_{s}},x_{\tilde{\chi}_j^0})
\Big\}\gamma^{\mu}\omega_-\nonumber\\&&\hspace{-0.5cm}
+\Big\{\frac{e^3}{128\pi^2s_{_W}c_{_W}}\Big(\frac{c_{_W}^2}{s_{_W}^2}(\frac{m_{l^I}}
{c_{\beta}m_{_W}})^2-\frac{4s_{_W}^2}{c_{_W}^2}\Big)\nonumber\\&&\hspace{-0.5cm}+\frac{e^3}{2s_{_W}c_{_W}}
\sum_{i,\beta=1}^6\sum_{j=1}^4(\mathcal{G})_{i\beta}
(\mathcal{C}^I)_{ij}(\mathcal{C}^I)^*_{\beta j}F_{1}(x_{\tilde{L}_{i}},x_{\tilde{L}_{\beta}},x_{\tilde{\chi}_j^0})
\nonumber\\&&\hspace{-0.5cm}
+\frac{e^3}{4s_{_W}c_{_W}}\sum_{s=1}^6\sum_{i,j=1}^4
(\mathcal{C}^I)^*_{si}(\mathcal{R})_{ji}(\mathcal{C}^I)_{sj}
F_{1}(x_{\tilde{\chi}_{i}^0},x_{\tilde{L}_{s}},x_{\tilde{\chi}_j^0})
\nonumber\\&&\hspace{-0.5cm}-\frac{e^3}{2s_{_W}c_{_W}}\sum_{s=1}^6\sum_{i,j=1}^4
(\mathcal{C}^I)^*_{si}(\mathcal{R})^*_{ji}
(\mathcal{C}^I)_{sj}\sqrt{x_{\tilde{\chi}_{i}^0}x_{\tilde{\chi}_{j}^0}}
F_2(x_{\tilde{\chi}_{i}^0},x_{\tilde{L}_{s}},x_{\tilde{\chi}_j^0})
\Big\}\gamma^{\mu}\omega_+.\label{Fllz}
\end{eqnarray}

To get Eq.(\ref{Fllz}), we use
the unitary character of the mixing matrixes $\mathcal{Z}_{\tilde{L}}, \mathcal{Z}_{N}$
for sleptons and neutralinoes. Additionally, we adopt the abbreviation notations $c_{_W}=\cos\theta_{_W},
s_{_W}=\sin\theta_{_W}, c_{_\beta}=\cos\beta, s_{_\beta}=\sin\beta$, where $\theta_{_W}$
is the Weinberg angle and $\tan\beta=v_2/v_1$ representing the ratio between the vacuum
expectation values of the two Higgs doublets. $x_i=m_i^2/\Lambda^2_{_{NP}}$ with $i$ denoting
the virtual particles in these one loop diagrams, and $\Lambda_{_{NP}}$ denotes
the new physic energy scale.
Here $\Delta_{UV}=1/\epsilon+\ln(4\pi x_{\mu})-\gamma_{_E},\;2\epsilon=4-D$, $\omega_-=(1-\gamma_5)/2,\;\omega_+=(1+\gamma_5)/2$ and the functions $F_1,F_2$ are shown as
\begin{eqnarray}
&&F_1(x,y,z)=\frac{1}{16 \pi ^2}\Big(1-\frac{x^2\ln x }{(y-x) (z-x)}-\frac{y^2
   \ln y}{(x-y) (z-y)}-\frac{z^2 \ln z}{(x-z)(y-z)}\Big),\\
&&F_2(x,y,z)=\frac{1}{16\pi^2}\Big(\frac{x\ln x}{(y-x)(z-x)}+\frac{y\ln y}{(x-y)(z-y)}+\frac{z\ln z}{(x-z)(y-z)}\Big).
\end{eqnarray}
The concrete forms of the vertex couplings used in  Eq.(\ref{Fllz}) reads as
\begin{eqnarray}
&&(\mathcal{C}^I)_{tj}=\frac{-\sqrt{2}}{c_{_W}}\mathcal{Z}_{\tilde{L}}^{(I+3)t}\mathcal{Z}_{_N}^{1j*}
-\frac{m_{l^I}\mathcal{Z}_{\tilde{L}}^{It}\mathcal{Z}_{_N}^{3j*}}{\sqrt{2}s_{_W}c_{\beta}m_{_W}},\nonumber\\
&&(\mathcal{D}^I)_{tj}=\frac{\mathcal{Z}_{\tilde{L}}^{It}}{c_{_W}}
(\!\mathcal{Z}_{_N}^{1j}s_{_W}\!+\!\mathcal{Z}_{_N}^{2j}c_{_W}\!)
\!-\!\frac{m_{l^I}\mathcal{Z}_{\tilde{L}}^{(I+3)t}
\mathcal{Z}_{_N}^{3j}}{c_{\beta}m_{_W}},\nonumber\\
&&(\mathcal{G})_{ts}=\frac{1}{2}\mathcal{Z}_{\tilde{L}}^{It}
\mathcal{Z}_{\tilde{L}}^{Is*}-s^2_{_W}\delta^{st},~~(\mathcal{R})_{k\alpha}=(\mathcal{Z}_N^{4k}\mathcal{Z}_N^{4\alpha *}-\mathcal{Z}_N^{3k}\mathcal{Z}_N^{3\alpha *}).
\end{eqnarray}

In the on-shell scheme, the counter term for the radiative correction to SM vertex
$Z\overline{l^I} l^I$ is shown here\cite{onshell1}.
\begin{eqnarray}
&&\delta V_{Z\overline{l^I} l^I}^{(OS),\mu}= -\frac{e}{2}\Big[\delta
Z_{AZ}-\frac{1}{2s_{_W}^3c_{_W}}\Big(\frac{\delta
m_{_Z}^2}{m_{_Z}^2}-\frac{\delta m_{_W}^2}{m_{_W}^2}\Big)-\frac{(2s_{_W}^2-1)}{s_{_W}c_{_W}} \Big({\delta e\over
e}+\frac{1}{2}\delta Z_{ZZ}\nonumber\\
&&+\delta
Z_L^{l}\Big)\Big]\gamma^\mu\omega_-
-{e\over2}\Big[\delta
Z_{AZ}-\frac{s_{_W}}{c_{_W}}\Big(2{\delta e\over e}
+\frac{1}{s_{_W}^2}(\frac{\delta m_{_Z}^2}{m_{_Z}^2}-\frac{\delta
m_{_W}^2}{m_{_W}^2}) +\delta Z_{ZZ}+2\delta
Z_R^{l}\Big)\Big]\gamma^\mu\omega_+. \label{onzll}
\end{eqnarray}
$\delta Z_{A Z}$ and $\delta e$ are the
renormalization constants for $\gamma Z$ mixing and charge respectively.
Only the sum of amplitude of the two triangle diagrams can be renormalized
in the on-shell scheme. That is to say, the divergent term of each diagram can
not be canceled by the counter term. Another character is that just the
lepton wave function renormalization constants $\delta
Z^l_L$ and $\delta Z^l_R$ are necessary to counteract the ultra-divergent terms.

The renormalization constants for the left- and right-handed lepton wave
functions are deduced from the lepton self-energy with virtual SUSY particles
($\tilde{L},\tilde{\chi}^0$).
\begin{eqnarray}
&&\delta Z_L^l=-\frac{e^2}{64\pi^2s_{_W}^2}\Big(\frac{1}{c_{_W}^2}+(\frac{m_{l^I}}
{c_{\beta}m_{_W}})^2\Big)\Delta_{UV}-\sum_{i=1}^6\sum_{j=1}^4\Big\{\frac{1}{2s^2_{_W}}|
(\mathcal{D}^I)_{ij}|^2F_4(x_{\tilde{L}_{i}},x_{\tilde{\chi}_j^0})\nonumber\\&&
+x_{l^I}\Big[\frac{1}{2s^2_{_W}}|(\mathcal{D}^I)_{ij}|^2+|(\mathcal{C}^I)_{ij}|^2
+\frac{\sqrt{2}}{s_{_W}}\mathbf{Re}[(\mathcal{C}^{I})^{\dag}_{ij}
(\mathcal{D}^{I})_{ij}]F_3(x_{\tilde{L}_{i}},x_{\tilde{\chi}_j^0})\Big]\Big\},\nonumber
\\&& \delta Z_R^l=-\frac{e^2}{32\pi^2}\Big(\frac{2}{c_{_W}^2}+(\frac{m_{l^I}}
{\sqrt{2}s_{_W}c_{\beta}m_{_W}})^2\Big)\Delta_{UV}
-\sum_{i=1}^6\sum_{j=1}^4\Big\{|(\mathcal{C}^I)_{ij}|^2F_4(x_{\tilde{L}_{i}},x_{\tilde{\chi}_j^0})
\nonumber\\&&+x_{l^I}\Big[\frac{1}{2s^2_{_W}}|(\mathcal{D}^I)_{ij}|^2+|(\mathcal{C}^I)_{ij}|^2
+\frac{\sqrt{2}}{s_{_W}}\mathbf{Re}[(\mathcal{C}^{I})^{\dag}_{ij}(\mathcal{D}^{I})_{ij}]\Big]
F_3(x_{\tilde{L}_{i}},x_{\tilde{\chi}_j^0})\Big\}\label{RllSL},
\end{eqnarray}
where the functions $F_3, F_4$ are shown as follows
\begin{eqnarray}
&&\hspace{-1.6cm}F_3(x,y)=\frac{x^2+2xy(\ln
y-\ln x)-y^2}{32\pi^2(x-y)^3},\nonumber\\&&\hspace{-1.6cm}
F_4(x,y)=\frac{(2y-x)(y-x+x\ln x)-y^2\ln y }{32\pi^2(x-y)^2}.\label{F34}
\end{eqnarray}
Considering Eqs.(\ref{onzll})(\ref{RllSL})(\ref{F34}), the needed counter terms
 for Diagram 1(a) and Diagram 1(b) are obtained in the on-shell scheme.

 \subsection{$W^+\overline{\nu^I}l^I$ vertex with virtual SUSY particles ($\tilde{L},\tilde{\nu},\tilde{\chi}^0,\tilde{\chi}^{\pm}$)}
The condition of $W^+\overline{\nu^I}l^I$ vertex is more complex
than that of $Z\overline{l^I}l^I$ vertex. The three triangle
diagrams(2(a), 2(b) and 2(c)) are all necessary and belong to one suit, where
the virtual SUSY particles are
$\tilde{L},\tilde{\nu},\tilde{\chi}^0,\tilde{\chi}^{\pm}$. We collect the counter terms of these three diagrams
in the zero-momentum scheme.
\begin{eqnarray}
&& \delta V_{W^+\overline{\nu^I}l^I}^{(ZM),\mu}
 =\frac{e^3}{\sqrt{2}s^3_{_W}}\Big\{\frac{1}{64\pi^2}\Big[\frac{1}{c_{_W}^2}+
 \Big(\frac{m_{l^I}}{c_\beta
 m_{_W}}\Big)^2\Big]\Delta_{UV}\nonumber\\&&
 +\frac{1}{128\pi^2}\Big[4-\frac{1}{c_{_W}^2}
+\Big(\frac{m_{l^I}}{c_\beta
m_{_W}}\Big)^2\Big]\nonumber\\&&
-\frac{1}{2c_{_W}}\sum_{i=1}^2\sum_{J=1}^3\sum_{j=1}^4
\Big((\zeta^I)^*_{Jj}(\mathcal{T})_{ji}(\mathcal{B}_i)^{IJ}
F_1\nonumber\\&&+2(\zeta^I)^*_{Jj}(\mathcal{Q})_{ji}
(\mathcal{B}_i)^{IJ}\sqrt{x_{\tilde{\chi}^-_{i}}x_{\tilde{\chi}_j^0}}
F_2\Big)(x_{\tilde{\chi}^-_{i}},x_{\tilde{\nu}_{J}},x_{\tilde{\chi}_j^0})\nonumber\\&&
+\frac{1}{4c_{_W}}\sum_{i=1}^6\sum_{J=1}^3\sum_{j=1}^4(\eta)^*_{iJ}(\zeta^I)^*_{Jj}(\mathcal{D}^I)_{ij}
F_{1}(x_{\tilde{L}_{i}},x_{\tilde{\nu}_{J}},x_{\tilde{\chi}_j^0})
\nonumber\\&&
+\frac{1}{2}\sum_{i=1}^2\sum_{s=1}^6\sum_{j=1}^4\Big(2(\mathcal{P}^I)_{si}(\mathcal{Q})^*_{ji}
(\mathcal{D}^I)_{sj}\sqrt{x_{\tilde{\chi}^-_{i}}x_{\tilde{\chi}_j^0}}
F_2\nonumber\\&&+(\mathcal{P}^I)_{si}(\mathcal{T})^*_{ji}(\mathcal{D}^I)_{sj}
F_1\Big)(x_{\tilde{\chi}^-_{i}},x_{\tilde{L}_{s}},x_{\tilde{\chi}_j^0})\Big\}\gamma^{\mu}\omega_-,
\end{eqnarray}
with the vertexes couplings
\begin{eqnarray}
&&(\mathcal{A}_i)^{IJ}={-m_{l^I}\mathcal{Z}^{-*}_{2i}
\mathcal{Z}^{IJ*}_{\tilde{\nu}}\over\sqrt{2}c_{_{\beta}}m_{_W}},~~
(\mathcal{B}_i)^{IJ}=\mathcal{Z}^{+}_{1i}\mathcal{Z}^{IJ*}_{\tilde{\nu}},\nonumber\\
&&(\mathcal{\eta})_{s_J}=\mathcal{Z}_{\tilde{\nu}}^{IJ}\mathcal{Z}_{\tilde{_L}}^{Is},~~
(\mathcal{Q})_{ji}=Z_N^{2j*}Z_{1i}^+-\frac{Z_N^{4j*}Z_{2i}^+}{\sqrt{2}},\nonumber\\&&
(\mathcal{T})_{ji}=Z_N^{2j}Z_{1i}^-+\frac{1}{\sqrt{2}}Z_N^{3j}Z_{2i}^{-*},\nonumber\\
&&(\mathcal{P}^I)_{si}=\frac{m_{l^I}}{\sqrt{2}c_{_{\beta}}M_{_W}}
Z_{\tilde{L}}^{(I+3)s*}Z_{2i}^{-*}-Z_{\tilde{L}}^{Is*}Z_{1i}^{-*},
\nonumber\\&&(\mathcal{\zeta}^{I})_{_Jj}=\mathcal{Z}_{\tilde{\nu}}^{IJ*}
(\mathcal{Z}_{_N}^{1j}s_{_W}-\mathcal{Z}_{_N}^{2j}c_{_W}).
\end{eqnarray}
In the on-shell scheme, the counter term formula for vertex $W^+\overline{\nu^I}l^I$ can be found in Ref\cite{onshell1}.
\begin{eqnarray}
&&\delta V_{W^+\overline{\nu^I}l^I}^{(OS),\mu}=
\frac{e}{2\sqrt{2}s_{_W}}\Big(\frac{\delta
m_{_Z}^2}{m_{_Z}^2}-\frac{\delta m_{_Z}^2 -\delta m_{_W}^2}{m_{_Z}^2-m_{_W}^2}+2\delta
e+\delta Z_{L}^{l}+\delta Z_{L}^{\nu}+\delta
Z_{WW}\Big)\gamma^\mu\omega_- \label{counter-os2},
\end{eqnarray}

where $\delta e$ is calculated from the virtual slepton
contribution. The virtual slepton and sneutrino produce the mass
renormalization constants $\delta m_{_Z}^2, \delta m_{_W}^2$ and $W$
wavefunction renormalization constant $\delta Z_{WW}$. The wave function renormalization
constants $\delta Z_L^{\nu}$ and
$\delta Z_L^l$ are deduced respectively from
the self-energies of neutrino and lepton with virtual SUSY
particles $[(\tilde{\nu},\tilde{\chi}^0),(\tilde{L},\tilde{\chi}^{\pm})]$ and
$[(\tilde{\nu},\tilde{\chi}^{\pm}),(\tilde{L},\tilde{\chi}^0)]$.

To cancel the UV-divergent terms for these diagrams in the on-shell scheme, all the
renormalization constants in Eq.(\ref{counter-os2}) must be taken
into account. Following the method in Refs\cite{onshell1,onshell2}, we obtain the needed
renormalization constants.
\begin{eqnarray}
&&\frac{\delta
m_{_Z}^2}{m_{_Z}^2}=\frac{e^2}{32\pi^2s_{_W}^2c_{_W}^2}(1-2s_{_W}^2)^2\Delta_{UV}
-\frac{e^2}{s_{_W}^2c_{_W}^2}\Big\{\frac{1}{4}\sum_{j=1}^3F_5(x_{\tilde{\nu}_{j}},x_{\tilde{\nu}_{j}})
\nonumber\\&&+\sum_{\alpha,\beta=1}^6|(\mathcal{G})_{\alpha\beta}|^2
F_5(x_{\tilde{L}_{\alpha}},x_{\tilde{L}_{\beta}})\Big\},\nonumber\\
&& \frac{\delta
m_{_W}^2}{m_{_Z}^2}=\frac{e^2c_{_W}^2}{32\pi^2s_{_W}^2}\Delta_{UV}
-\frac{e^2c_{_W}^2}{2s_{_W}^2}\sum_{i=1}^6\sum_{\alpha=1}^3
|(\mathcal{\eta})_{i\alpha}|^2F_5(x_{\tilde{\nu}_{\alpha}},x_{\tilde{L}_{i}}),\nonumber\\&&
\delta
Z_{WW}=-\frac{e^2}{32\pi^2s_{_W}^2}\Delta_{UV}+\frac{e^2}{2s_{_W}^2}\sum_{i=1}^6\sum_{\alpha=1}^3
|(\mathcal{\eta})_{i\alpha}|^2F_5(x_{\tilde{\nu}_{\alpha}},x_{\tilde{L}_{i}}),\label{ww}\nonumber\\
&&\delta
e=\frac{e^2}{8\pi^2}\Delta_{UV}-e^2\sum_{i=1}^6F_5(x_{\tilde{L}_{i}},x_{\tilde{L}_{i}}),\nonumber\\
&& \delta
Z_L^{\nu}=\frac{-e^2}{32\pi^2s_{_W}^2}\Big(\frac{1}{2c_{_W}^2}+1
+(\frac{m_{l^I}}{\sqrt{2}c_{\beta}m_{_W}})^2\Big)\Delta_{UV}\nonumber\\&&
-\frac{e^2}{2s^2_{_W}c^2_{_W}}\sum_{i=1}^4\sum_{\alpha=1}^3|(\zeta^I)_{\alpha i}|^2
F_4(x_{\tilde{\nu}_{\alpha}},x_{\chi_{i}^0})-\frac{e^2}{s^2_{_W}}\sum_{i=1}^4\sum_{\alpha=1}^6|(\mathcal{P}^I)_{\alpha i}|^2
F_4(x_{\tilde{L}_{\alpha}},x_{\chi_{i}^0}),\nonumber\\
&&\delta
Z_L^l=-\frac{e^2}{32\pi^2s_{_W}^2}\Big(\frac{1}{2c_{_W}^2}+1
+(\frac{m_{l^I}}{\sqrt{2}c_{\beta}m_{_W}})^2\Big)\Delta_{UV}\nonumber\\&&
-\frac{e^2}{s_{_W}^2}\sum_{\alpha=1}^3\sum_{i=1}^2\Big\{|(\mathcal{B}_i)^{I\alpha}|^2
F_4+x_{l^I}\Big[|(\mathcal{B}_i)^{I\alpha}|^2\nonumber\\&&+|(\mathcal{A}_i)^{I\alpha}|^2+
2\mathbf{Re}[(\mathcal{A}_i^{\dag})^{I\alpha}(\mathcal{B}_i)^{I\alpha}]\Big]
F_3\Big\}(x_{\tilde{\nu}_{\alpha}},x_{\tilde{\chi}_i^-})\nonumber\\&&
-e^2\sum_{j=1}^4\sum_{i=1}^6\Big\{x_{l^I}\!\Big[\frac{|(\mathcal{D}^I)_{ij}|^2}{2s^2_{_W}}\!
+\!\frac{\sqrt{2}}{s_{_W}}\mathbf{Re}[(\mathcal{C}^I)^{\dag}_{ij}
(\mathcal{D}^I)_{ij}]\nonumber\\&&+|(\mathcal{C}^I)_{ij}|^2\Big]F_3
+\frac{1}{2s^2_{_W}}|(\mathcal{D}^I)_{ij}|^2F_4\Big\}(x_{\tilde{L}_{i}},x_{\tilde{\chi}_j^0}).\label{VWLos}
\end{eqnarray}
The function $F_5$ is
\begin{eqnarray}
&&\!\!F_5(x,y)=\frac{1}{288\pi^2(x-y)^3}[6(x-3y)x^2\ln
x\!+\!6(3x\nonumber\\&&\!\!-y)y^2\ln y-(x-y)(5x^2-22xy+5y^2)].\label{F5}
\end{eqnarray}
From Eqs.(\ref{counter-os2})(\ref{VWLos})(\ref{F5}), we get the counter terms for the three diagrams(2(a),2(b) and 2(c)). The
renormalization constants in Eq.(\ref{counter-os2}) are all
necessary at this place, which is different from the condition of
$Z\overline{l^I}l^I$ vertex.
 \section{Renormalization of $\tilde{L}^*_i\overline{\chi^0_{\alpha}}l^I$ vertex with virtual photon}
 In order to further research the on-shell renormalization scheme in MSSM,
  we study the vertex $\tilde{L}^*_i\overline{\chi^0_{\alpha}}l^I$ at
 one-loop order in this section. The studied triangle
diagram is Diagram 3 with virtual photon, which is the simplest instance.
Diagram 3 belongs to electromagnetic interaction, and can be treated separately without
considering the diagrams with virtual W and Z.
The counter term for this diagram in the zero-momentum scheme is
\begin{eqnarray}
&&\delta V_{\tilde{L}^*_{i}\overline{\chi^0_{\alpha}}l^I}^{(ZM)}(\gamma)=
\frac{e^3}{16\pi^2}\Big\{\frac{(\mathcal{D}^I_{i\alpha})}{\sqrt{2}s_{_W}}\omega_-\!
+\!(\mathcal{C}^I_{i\alpha})\omega_+\Big\}\Delta_{UV}\nonumber\\&&
+e^3F_{1}(x_{\tilde{L}_i},0,x_{l^I})\Big(\frac{(\mathcal{D}^I_{i\alpha})}{\sqrt{2}s_{_W}}\omega_-
+(\mathcal{C}^I_{i\alpha})\omega_+\Big)\label{zmsusyg}.
\end{eqnarray}
 In the on-shell scheme the counter term formula of the vertex $\tilde{L}^*_i\overline{\chi^0_{\alpha}}l^I$ is complicated.
  Following the idea of SM on-shell scheme, we show
 the formula here\cite{feng2}, where the counter term is determined by the on-shell condition.
\begin{eqnarray}
&&\delta V_{\tilde{L}^*_{i}\overline{\chi^0_{\alpha}}l^I}^{(OS)}\!=\!
{e\over\sqrt{2}s_{_W}c_{_W}}\Big\{\Big[\Big(\frac{\delta
e}{e}\delta_{IJ}+{(\delta
Z_L^l)_{JI}\over2}\Big)\delta_{ij}\delta_{\alpha\beta}+{(\delta
Z^{\dag}_{\tilde{L}})_{ij}\over2}\delta_{IJ}\delta_{\alpha\beta}
\nonumber\\
&& +{(\delta
Z_{\chi^0})_{\beta\alpha}\over2}\delta_{IJ}\delta_{ij}\Big]
(Z^{\dag}_{\tilde{L}})_{jJ}\Big(Z_N^{1\beta}s_{_W}
+Z_N^{2\beta}c_{_W}\Big) -{s_{_W}\over c_{_W}}\delta
c_{_W}(Z^{\dag}_{\tilde{L}})_{jJ}Z_N^{1\beta}\delta_{ij}\delta_{IJ}\delta_{\alpha\beta}\nonumber\\
&&-{c_{_W}\over s_{_W}}\delta
s_{_W}(Z^{\dag}_{\tilde{L}})_{jJ}Z_N^{2\beta}\delta_{ij}\delta_{IJ}\delta_{\alpha\beta}
 -{m_{l^J}c_{_W}\over
m_{_W}c_{\beta}}\Big[\Big(\frac{\delta e}{e}+\frac{\delta
m_{l^J}}{m_{l^J}} +{\delta m_{_W}\over m_{_W}}-{\delta s_{_W}\over
s_{_W}}-{\delta c_{\beta}\over c_{\beta}}\Big)
\delta_{IJ}\delta_{ij}\delta_{\alpha\beta}
\nonumber\\&&+{1\over2}(\delta
Z_{\tilde{L}}^{\dag})_{ij}\delta_{IJ}\delta_{\alpha\beta}
+{1\over2}(\delta
Z_{\tilde{\chi}^0})_{\beta\alpha}\delta_{IJ}\delta_{ij}
+{1\over2}(\delta Z^l_L)_{JI}\delta_{\alpha\beta}\delta_{ij}\Big]
(Z^{\dag}_{\tilde{L}})_{j(3+J)}Z_N^{3\beta}\Big\}\omega_-\nonumber\\&&
 +{\sqrt{2}e\over c_{_W}}\Big\{-\Big[\Big({\delta
e\over e}-{\delta c_{_W}\over c_{_W}}\Big)
\delta_{IJ}\delta_{ij}\delta_{\alpha\beta} +{1\over2}(\delta
Z_{\tilde{L}}^{\dag})_{ij}\delta_{IJ}\delta_{\alpha\beta}
+{1\over2}(\delta Z^*_{\tilde{\chi}^0})_{\beta\alpha}\delta_{IJ}\delta_{ij}
\nonumber\\&&+{1\over2}(\delta
Z^l_R)_{JI}\delta_{\alpha\beta}\delta_{ij}\Big](Z^{\dag}_{\tilde{L}})_{j(3+J)}Z_N^{1\beta
*} +{m_{l^J}c_{_W}\over2m_{_W}s_{_W}c_{\beta}}\Big[\frac{\delta
Z_{\tilde{L}}^{\dag})_{ij}}{2}(\delta_{IJ}\delta_{\alpha\beta}
+\frac{(\delta
Z^*_{\tilde{\chi}^0})_{\beta\alpha}}{2}\delta_{IJ}\delta_{ij}\nonumber\\
&&+\Big({\delta
e\over e}+{\delta m_{l^J}\over m_{l^J}} +{\delta m_{_W}\over m_{_W}}
-\frac{\delta s_{_W}}{s_{_W}}-\frac{\delta
c_{\beta}}{c_{\beta}}\Big)\delta_{IJ}\delta_{ij}\delta_{\alpha\beta}
+\frac{1}{2}(\delta
Z^l_R)_{JI}\delta_{\alpha\beta}\delta_{ij}\Big]
(Z^{\dag}_{\tilde{L}})_{jJ}Z_N^{3\beta*}\Big\}\omega_+\;.
\label{counter-os1}
\end{eqnarray}
 $\delta Z^l_{L,R},\delta Z_{\tilde{\nu}},\delta
Z_{\tilde{L}},\delta Z_{\tilde{\chi}^{-}}$ and $\delta
Z_{\tilde{\chi}^{0}}$ are the renormalization constants of wave functions for
leptons and SUSY particles. The other renormalization constants come from
the vertex coupling renormalization.

After tedious calculation and various compounding of renormalization constants, we
find only the wave function renormalization constant of slepton
$(\delta Z_{\tilde{L}})_{ij}$ is essential. That is to
say, just the renormalization constant $(\delta Z_{\tilde{L}})_{ij}$ can cancel the UV-divergent term.
The wave function renormalization constant
$(\delta Z_{\tilde{L}})_{ij}$ is collected in the as follows.
\begin{eqnarray}
&&\hspace{-1.0cm}F_6(x,y)=\frac{1}{32\pi^2(y-x)^3}[(y-x)(6x^2-7xy+3y^2)\nonumber\\&&
\hspace{-1.0cm}+2x(2x^2-2xy+y^2)\ln
x-2y(4x^2-5xy+2y^2)\ln y],
\nonumber\\&&\hspace{-1.0cm}(\delta Z^{\gamma}_{\tilde{L}})_{ij}=\frac{e^2}{8\pi^2}\Delta_{UV}\delta^{ij}
+e^2F_6(x_{\tilde{L}_i},0)\delta^{ij}.\label{gamma}
\end{eqnarray}
$(\delta Z_{\tilde{L}})_{ij}^{\gamma}$ in Eq.(\ref{gamma}) is
obtained from the self-energy of slepton with the virtual
photon and slepton.
In our calculation, Eq.(\ref{counter-os1}) is predigested as
\begin{eqnarray}
&&\hspace{-1.6cm}\delta V_{\tilde{L}^*_{i}\overline{\chi^0_{\alpha}}l^I}^{(OS)}(\gamma)={1\over2}(\delta
Z^{\gamma}_{\tilde{L}})^{\dag}_{ij}\Big[(\mathcal{D}^I_{j\alpha})\omega_-+(\mathcal{C}^I_{j\alpha})\omega_+\Big].\label{jianhua}
\end{eqnarray}
Combining the formulas (\ref{gamma}) and (\ref{jianhua}), Diagram 3 can be
renormalized successfully in the on-shell scheme. Up to now, we have got the counter terms for the vertexes $(Z\overline{l^I}l^I , W^+\overline{\nu^I}l^I )$ and $\tilde{L}^*_s\overline{\chi^0_j}l^I$ in both the zero-momentum and on-shell schemes.
\section{The decoupling behavior}
In this section, we discuss the decoupling behavior of renormalized results in the two schemes.
It is easy to prove that the renormalized results in the zero-momentum scheme are decoupled.
Adopting the on-shell
scheme, we must get decoupled renormalized results, if the renormalized
results can not go to infinity with the incessant enlarging
SUSY particle masses. To obtain the decoupling behavior of renormalized results in the on-shell scheme, we suppose all SUSY particle masses are the same and much heavier than the masses of SM particles.
Compared with the decoupling character of zero-momentum counter terms, the decoupling behavior of counter terms in the on-shell scheme is obvious.
\subsection{SM vertex $(Z\overline{l^I}l^I,
W^+\overline{\nu^I}l^I )$ }
To obtain the decoupling behavior of the counter terms for the vertex $Z\overline{l^I}l^I
$ in the zero-momentum scheme, we show the decoupling approximation of the functions $F_1$and $F_2$.
The variables $x,y,z$ in $F_1(x,y,z)$ are all symmetrical, and three conditions are considered here.
\begin{eqnarray}
&&\!\!F_1(x,y,z)=\Bigg\{
\begin{array}{l}
-\frac{\ln x}{16\pi^2}-\frac{1}{32\pi^2}, ~~~~~(x=y=z) \\
 -\frac{\ln x}{16\pi^2}+\dots~~~~~~~~(x=y \gg z)\\
\frac{1-\ln x}{16 \pi
   ^2}+\dots~~~~~~~~~(x \gg y,z)
\end{array}\nonumber
\\&&\!\!F_2(x,y,z)=\frac{1}{32\pi^2x},~(x=y=z).\label{tf12}
\end{eqnarray}
With Eqs.(\ref{Fllz}) and (\ref{tf12}), the decoupling behavior of Eq.(\ref{Fllz}) reads
\begin{eqnarray}
&&\hspace{-0.8cm}\delta V_{Z\overline{l^I} l^I}^{(ZM),\mu}\!\sim\!\frac{e^3}{64\pi^2s_{_W}c_{_W}}
\Big\{\!\frac{1\!-\!2s_{_W}^2}{2s_{_W}^2}\Big[\frac{1}{c_{_W}^2}\!
+\!\Big(\frac{m_{l^I}}{c_{\beta}
M_W}\Big)^2\Big]\gamma^{\mu}\omega_-\nonumber\\&&\hspace{-0.8cm}-\Big[\frac{4s_{_W}^2}{c_{_W}^2} +\Big(\frac{m_e^I}{c_{\beta}
M_W}\Big)^2
\Big]\gamma^{\mu}\omega_+\Big\}(\Delta_{UV}-\ln x_{_M})+\dots,\label{tzmzll}
\end{eqnarray}
where the dots denote the terms that are finite, even when the SUSY particle masses turn to infinity.
$x_{_M}=M^2/\Lambda^2_{_{NP}}$ with $M$ representing the SUSY particle mass.
  In the same way, we deduce the decoupling behavior of the counter terms
  for the vertex $Z\overline{l^I}l^I$ in the on-shell scheme.
\begin{eqnarray}
&&\delta V_{Z\overline{l^I} l^I}^{(OS),\mu}\!\sim\!\frac{e^3}{64\pi^2s_{_W}c_{_W}}
\Big\{\!\frac{1\!-\!2s_{_W}^2}{2s_{_W}^2}\Big[\frac{1}{c_{_W}^2}
\!+\!\Big(\frac{m_{l^I}}{c_{\beta}
M_W}\Big)^2\Big]\gamma^{\mu}\omega_-\nonumber\\&&-\Big[\frac{4s_{_W}^2}{c_{_W}^2} +\Big(\frac{m_{l^I}}{c_{\beta}
m_{_W}}\Big)^2
\Big]\gamma^{\mu}\omega_+\Big\}(\Delta_{UV}-\ln x_{_M})+\dots\label{toszll}\\
&&F_3(x,y)=\frac{1}{96\pi^2x},~(x=y);~~
F_4(x,y)=-\frac{\ln x}{32\pi^2}+\frac{1}{64\pi^2},~(x=y).
\end{eqnarray}

It is satisfactory that the infinite terms and undecoupled large logarithm terms in Eqs.(\ref{tzmzll}) and (\ref{toszll}) are the same.
Though the finite terms represented by dots in Eqs.(\ref{tzmzll}) and (\ref{toszll}) are different, the renormalized results in both schemes are decoupled, because the zero-momentum scheme can guarantee the decoupled renormalized results.

 Using the unitary character of the mixing matrixes we obtain the expectant
 results for the counter terms of the vertex $W^+\overline{\nu^I}l^I$ in both schemes.
\begin{eqnarray}
 &&\delta V_{W^+\overline{\nu^I}l^I}^{(ZM),\mu}
 \sim\Big\{\frac{e^3}{\sqrt{2}s^3_{_W}}\frac{1}{64\pi^2}\Big[\Big(\frac{m_{l^I}}{c_\beta
 M_W}\Big)^2+\frac{1}{c_{_W}^2}
 \Big](\Delta_{UV}-\ln x_{_M})\Big\}\gamma^{\mu}\omega_-+\dots,\label{tzmwnl}\\
 &&\delta V_{W^+\overline{\nu^I}l^I}^{(OS),\mu}
  \sim \Big\{\frac{e^3}{\sqrt{2}s^3_{_W}}\frac{1}{64\pi^2}\Big[\Big(\frac{m_{l^I}}{c_\beta
 M_W}\Big)^2+\frac{1}{c_{_W}^2}
 \Big](\Delta_{UV}-\ln x_{_M})\Big\}\gamma^{\mu}\omega_-+\dots,\label{toswnl}\\
 &&F_5(x,y)=\frac{\ln x}{48\pi^2},~(x=y)\nonumber.
\end{eqnarray}
The two counter terms in Eqs.(\ref{tzmwnl}) and (\ref{toswnl}) can both eliminate completely
$\Delta_{UV}$ and $\ln x_{_M}$ terms produced from the three triangle diagrams(2(a), 2(b) and 2(c)).

\subsection{The MSSM vertex $\tilde{L}^*_i\overline{\chi^0_{\alpha}}l^I$ }
For the MSSM vertex $\tilde{L}^*_i\overline{\chi^0_{\alpha}}l^I$, the decoupling behavior of counter term is discussed here.
Assuming SUSY particles are very heavy, the approximate results of Eq.(\ref{zmsusyg}) deduced from virtual photon are shown as
\begin{eqnarray}
&&\delta V_{\tilde{L}^*_{i}\overline{\chi^0_{\alpha}}l^I}^{(ZM)}(\gamma)\sim
\frac{e^3}{16\pi^2}\Big\{\frac{1}{\sqrt{2}s_{_W}}(\mathcal{D}^I_{i\alpha})\omega_-
+(\mathcal{C}^I_{i\alpha})\omega_+\Big\}(\Delta_{UV}-\ln x_{_M})+\dots\nonumber
\\&&
F_6(x,y)=-\frac{\ln x}{8\pi ^2}+\frac{3}{16 \pi ^2}+\dots,~~(x\gg y).\label{zmgamma}
\end{eqnarray}

The decoupling behavior of the counter term Eq.(\ref{jianhua}) in the on-shell scheme is the same as
that of Eq.(\ref{zmgamma}) for $\Delta_{UV}$ and $\ln x_{_M}$. In this way, we find that the renormalized results are decoupled not only in the zero-momentum scheme
but also in the on-shell scheme.

\section{Discussion and conclusion}
Up to now there have been several renormalization schemes for
renormalizable theories. The on-shell renormalization scheme is approbated broadly
for electroweak theory in SM, and it is well studied by theorists.
For the model including new physics beyond SM,
the on-shell renormalization scheme has mist
to clear. MSSM is considered the most potent candidate in the new
models, which has attracted much attention from many people
for about twenty years. In the frame work of MSSM, some processes
are calculated with the on-shell renormalization scheme. However, a
consummate on-shell renormalization scheme for MSSM is still absent.

To explore the perfect on-shell renormalization scheme, at one-loop order we study two SM
vertexes$(Z\overline{l^I}l^I , W^+\overline{\nu^I}l^I )$ and one MSSM
vertex $\tilde{L}^*_i\overline{\chi^0_{\alpha}}l^I$ in
the zero-momentum scheme and the on-shell scheme. In the zero-momentum scheme, each divergent diagram
has its own counter term, and has nothing to do with other diagrams. Another important peculiarity
is that the renormalized result is absolutely decoupled.

In the on-shell scheme, the counter term formulas for the SM vertexes in MSSM
and in SM are similar. Almost all the renormalization constants
are deduced from the one-loop self-energies of the corresponding
particles. In SM, all the renormalization constants in the counter
term must be taken into account. At the same time, in MSSM we
can not always renormalize one triangle diagram by the counter term made
up of renormalization constants. After careful study, both
characters of the on-shell scheme are discovered. One character is that
all the triangle diagrams belonging to one type for a vertex are essential. Only
the sum of the amplitude can be renormalized completely.
The other character is that not all the renormalization constants are
always necessary. Which renormalization constant must be considered
 lies on the idiographic condition.

This work shows that for the SM vertex $Z\overline{l^I}l^I$ the lepton wave
function renormalization constants $\delta Z_L^{l},\delta Z_R^{l}$
are requisite to obtain the needed counter term. However, the condition of
the vertex $W^+\overline{\nu^I}l^I$ is dissimilar. To gain the final
finite results, we have to calculate all the renormalization
constants in the counter term formula. For the MSSM vertex, the foregoing
experience is of value of reference. The on-shell scheme for
the MSSM vertex $\tilde{L}^*_i\overline{\chi^0_{\alpha}}l^I$ shows the property,
i.e. just the wave function renormalization constant for the
relevant slepton $(\tilde{L})$ is enough for completing
the on-shell scheme.

In the two renormalization schemes, we study the decoupling behavior for
the counter terms of these vertexes. Obviously, the counter terms
obtained in the two renormalization schemes
have the same characters for the infinite and large logarithm terms,
when the SUSY particle masses are equal and very heavy.
Because the renormalized results in the zero-momentum scheme are
 decoupled, the on-shell renormalization scheme can also give decoupled renormalized results.

 There are a great deal of vertexes in MSSM, so it is hard to make
 one-loop on-shell renormalization for all of them. The studied vertexes in
 this work are representative, which can be helpful to upbuild a consistent on-shell
 renormalization scheme in MSSM. Though there are a number of
 questions to deal with, one can be convinced that a perfect on-shell
 renormalization scheme can be found in the future. This text is also propitious
 to study the on-shell renormalization scheme in other models, even the
 model is more complex than MSSM.


\begin{acknowledgments}
The work has been supported by the National Natural Science Foundation
of China (NNSFC) with Grant No. 11275036 and 11047002.
One of authors (SMZ) is also supported by the Fund of Natural
 Science Foundation of Hebei Province(A2011201118) and
 Natural Science Fund of Hebei University (2011JQ05 and 2012-242).
\end{acknowledgments}

\end{document}